\documentclass[prl,twocolumn,showpacs]{revtex4}

\usepackage{graphicx}
\usepackage{dcolumn}
\usepackage{bm}

\begin{document}


\title{Phase-matched four wave mixing and quantum beam splitting of matter waves
in a periodic potential}

\author{Karen Marie Hilligs\o e}
\email{kmh@phys.au.dk}
\author{Klaus M\o lmer}
\affiliation{Danish National Research Foundation Center for Quantum Optics,
Department of Physics and Astronomy, University of Aarhus, DK-8000 \AA rhus C,
Denmark}


\begin{abstract}
We show that the dispersion properties imposed by an external
periodic potential ensure  both energy and quasi-momentum
conservation such that correlated pairs of atoms can be generated
by four wave mixing from a Bose-Einstein condensate moving in an
optical lattice potential. In our numerical solution of the
Gross-Pitaevskii equation, a condensate with initial
quasi-momentum $k_0$ is transferred almost completely ( $>$ 95\%)
into a pair of correlated atomic components with quasi-momenta
$k_1$ and $k_2$, if the system is seeded with a smaller number of
atoms with the appropriate quasi-momentum $k_1$.
\end{abstract}

\pacs{03.75.Kk, 03.75.Lm, 05.45.-a}
\maketitle

Bose-Einstein condensates in optical lattices provide flexible
systems for studying the behavior of coherent matter in periodic
potentials. Considerable attention is given to studies in regimes
far from the region of validity of mean-field analysis and the
Gross-Pitaevskii equation \cite{bloch04a}, but also the highly
non-trivial mean-field dynamics has been and continues to be
subject to theoretical and experimental investigations
\cite{wu03a,morsch02a,hilligsoe02a}.

The process we wish to consider is a four wave mixing (FWM)
process, which transfers pairs of atoms coherently from an initial
momentum state $k_0$ to new states with momenta $k_1$ and $k_2$.
We consider a Bose-Einstein condensate in a quasi-1D geometry and
we will consider only the longitudinal dynamics of the condensate.
This geometry is relevant, e.g., for atomic wave guides and atom
interferometers based on atom chips \cite{hommelhoff04a}.

In Refs.~\cite{zhang03a,bouchoule03a}, it was shown that nonlinear
interaction originating from the s-wave scattering between atoms
leads to depletion of the condensate and emission of pairs of
atoms at other momenta when a continuous matter wave beam passes
through a finite region with enhanced interactions. For a larger
condensate, however, the process will not be effective unless it
conserves both energy and momentum, i.e., the waves must be
phase-matched over the extent of the sample. We shall show how the
characteristic energy band structure in a one-dimensional optical
lattice can be used to ensure both energy and quasi-momentum
conservation, i.e., phase-matching of the FWM process.

Our tailoring of the dispersion properties of matter waves by an
external potential is inspired by approaches to non-linear optics,
which employ various means to ensure phase-matching, e.g., of the
FWM process \cite{agrawal01a,hilligsoe04a,andersen04a}. A similar
phase-matched FWM process has been used to explain giant
amplification from semiconductor microcavities, where the
polariton dispersion properties can be controlled by the strong
photon-exciton coupling \cite{savvidis00a}. We also note that a
recent analysis \cite{yulin03a} of the break-up of a bright matter
wave soliton was analyzed in terms of dispersion and
phase-matching. Phase-matched FWM has been realized in collisions
of two condensates in two dimensions
\cite{vogels02a,deng99a,trippenbach98a}, but in the present paper
we show that the process can take place with atomic motion along a
single direction, for example inside an atomic waveguide.

\begin{figure}
\includegraphics[scale=0.89]{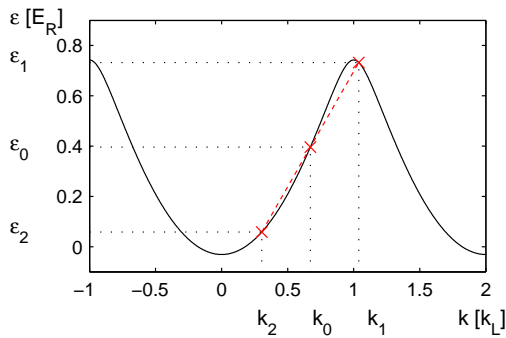}
\caption{\label{fig:1} Band structure for atomic motion in the
periodic potential Eq.~(\ref{Vz}). Quasi-momentum conservation and
energy conservation is fulfilled in the crosses where two atoms
with momentum $k_0$ collide and separate at momenta $k_1$ and
$k_2$ illustrated in the figure. }\label{fig1}
\end{figure}

The basic idea of our proposal is illustrated in Fig.~\ref{fig1}.
In a periodic potential $V(z)$, the energy spectrum constitutes a
band structure, and the figure shows the lowest energy band for
the corresponding linear Schr\"{o}dinger equation. When two atoms
with momentum $k_0$ collide and leave with momenta $k_1$ and $k_2$
momentum conservation requires
\begin{equation}\label{momentumconservation}
2k_0=k_1+k_2~modulo~Q,
\end{equation}
where $Q$ is a reciprocal lattice vector. In the periodic
potential the energy does not vary quadratically with the wave
number, and as indicated by example in Fig.~\ref{fig1}, it is
possible to identify sets of wave numbers with conservation of the
total energy
\begin{equation}\label{energyconservation}
2\varepsilon_0=\varepsilon_1+\varepsilon_2.
\end{equation}

To investigate the effectiveness of this degenerate FWM process,
we have performed a mean-field analysis of the dynamics of the
condensate based on the one-dimensional Gross-Pitaevskii equation
\begin{eqnarray}\label{1DGPE}
i\hbar \frac{d\Psi}{dt} =
\Big(-\frac{\hbar^2}{2m}\frac{d^2}{dz^2}+V(z)+\gamma|\Psi|^2\Big) \Psi,
\end{eqnarray}
where the periodic potential $V(z)$ is given by
\begin{eqnarray}\label{Vz}
V(z)=-\beta E_R cos(2k_Lz),
\end{eqnarray}
where $E_R\equiv \hbar^2k_L^2/2m$ is the recoil energy. The periodic potential can
be generated with a standing wave of a laser with wavelength $\lambda=2\pi/k_L$.
The factor
\begin{eqnarray}\label{gamma}
\gamma=gN/A_{\bot}
\end{eqnarray}
describes the strength of the nonlinearity, where $N$ is the total
number of atoms in the condensate, $A_{\bot}$ is the area of the
transverse ground state, and $g$ relates to the s-wave scattering
length $a_s$ and the mass $m$ of the atoms as
$g=4\pi\hbar^2a_s/m$. In our calculation presented below we assume
$\beta=1/2$ and $\gamma=40.8E_{R}/k_L$ corresponding to $N=100000$
$^{87}Rb$ atoms confined to a transverse area of
$A_{\bot}=42\mu$m$^2$ and a grid of 512 periods of the potential.
The effective area $A_\perp
=\frac{2\pi\hbar}{m\omega_{\perp}}\sqrt{1+2 a_s N/L}$
 results from a gaussian variational ansatz \cite{salasnich02a,modugno04a},
to the radial distribution in a harmonic potential with
$\omega_{\perp}=2\pi\times 44$ Hz and a constant longitudinal
density $N/L$.

An eigenstate $\Psi_0$ of the condensate with quasi-momentum $k_0$
is found using the method of steepest descent in imaginary time,
assuming a solution according to Bloch's theorem on a single
period of the lattice potential. Subsequently, a seed at variable
$k_1$ is applied giving the following initial wave function
\begin{eqnarray}\label{psis}
\Psi_{init}(z)=\frac{1}{\sqrt{1+\alpha^2}}\bigg[1+\alpha
e^{i(k_1-k_0)z}\bigg]\Psi_0(z),
\end{eqnarray}
where $\alpha=0.1$ has been used in our calculations with the
Gross-Pitaevskii equation. This wave function does not fulfill
Bloch's theorem, and we hence restrict ourselves to values of
$k_0$ and $k_1$ with interference patterns which are periodic on
an extended grid of 512 periods of the potential. To test the
importance of phase-matching in the FWM process we propagate the
wave function $\Psi_{init}$ on this grid, with different values
for the seeded momentum component $k_1$. As a function of time, we
can observe the evolution of the Gross-Pitaevskii wave function
and build-up of amplitude at different momenta. We are
particularly interested in the quasi-momentum regions around
$k_0$, $k_1$ and $k_2=2k_0-k_1$. Let $\psi(k,t)$ denote the
Fourier transform of the time-dependent Gross-Pitaevskii
wave-function $\Psi(z,t)$. The distribution in momentum space
folded into the single Brillouin zone from $k=0$ to $Q=2k_L$ is
given by the following expression
\begin{equation}
P_{k}(t)=\sum_n\int_{k-\Delta k/2}^{k+\Delta
k/2}|\psi(k+nQ,t)|^2dk,
\end{equation}
where we have introduced a sampling over a narrow momentum window with
$\Delta k=k_L/32 $.

\begin{figure}
\includegraphics[scale=0.89]{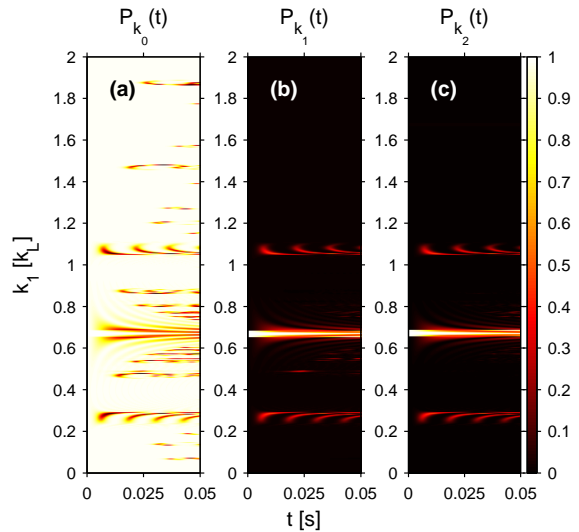}
\caption{ Population of different momentum components (a):
$P_{k_0}(t)$, (b): $P_{k_1}(t)$ and (c): $P_{k_2}(t)$ as a
function of time and as function of the seeding wave vector $k_1$.
The calculations are performed with a potential modulation
$\beta=1/2$, $N=100000$ atoms, and an initial wave vector of
$k_0=0.672k_L$. When ($k_0,k_1,k_2$) fulfill the phase-matching
conditions in Eq.~(\ref{momentumconservation}) and
(\ref{energyconservation}), which is the case for $k_1=0.289 k_L,\
1.055k_L$, the condensate originally having wave vector $k_0$ is
efficiently transferred into a set of atomic clouds with wave
vectors $k_1$ and $k_2$.}\label{fig2}
\end{figure}

Fig.~\ref{fig2}(a) shows the part of the condensate, $P_{k_0}(t)$,
remaining at the initial quasi-momentum $k_0$ when the condensate
is seeded with different values of $k_1$. The most important
features in the figure occur when the condensate is seeded with
$k_1=1.055k_L$ and $k_1=0.289k_L$. The original condensate
fraction at $k_0$ is almost completely depleted, and strong growth
of the population of the seeded momentum state is shown in
$P_{k_1}(t)$ in part b, accompanied by simultaneous growth in the
phase-matched component $k_2=2k_0-k_1$, shown as $P_{k_2}(t)$ in
part c of the figure. Comparing the set
$(k_0,k_1,k_2)=(0.672k_L,1.055k_L,0.289k_L)$ with the set of
phase-matched wave vectors in Fig.~\ref{fig1}, we find extremely
good agreement and we conclude that the structure at
$k_1=1.055k_L$ is indeed due to the phase-matched FWM process.

The remaining structures in Fig.~\ref{fig2} are less prominent but
for instance the structure in Fig.~\ref{fig2}(b) after 25 msec at
$k'_1=0.492k_L$ can be identified as a double FWM process with the
following steps: $2k_0\rightarrow k'_1+k'_2$ and
$k_0+k'_2\rightarrow k'_1+ k'_3$, where $k'_3=k'_2+(k_0-k'_1)$.
These steps do not conserve energy but the resulting six wave
mixing process ($3k_0\rightarrow2k'_1+k'_3$) is resonant. A closer
examination indeed confirms that the $k'_1$ component is twice as
populated as the $k'_3$ population in the region of the bright
spot on the figure. It is again the phase-matching that is
responsible for the significance of this amazing process.
\begin{figure}
\includegraphics[scale=0.89]{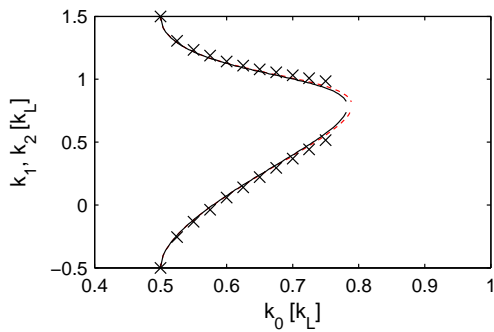}
\caption{Identification of momentum states $(k_0,k_1,k_2)$
fulfilling the phase-matching condition. Crosses: sets of
phase-matched values $(k_0, k_1, k_2)$ identified by numerical
solution of the Gross-Pitaevskii equation (\ref{1DGPE}). Full
line: phase-matching condition derived from band structure based
on simple linear Schr\"{o}dinger equation. Dashed line:
phase-matching condition when the effects of interactions have
been included as in Eq.~(\ref{popov}). Although the simple band
structure calculations are not exact theories for the
phase-matching condition there is still excellent agreement with
the full mean-field calculation.}\label{fig3}
\end{figure}

We have performed calculations with the Gross-Pitaevskii equation
(\ref{1DGPE}) for various $k_0$ and identified the phase-matched
sets of wave vectors $(k_0,k_1,k_2)$, plotted as crosses in
Fig.~\ref{fig3}. To quantitatively understand the occurring sets
of wave vectors we performed a simple band structure calculation
based on a linear Schr\"{o}dinger equation and extracted the sets
of wave vectors fulfilling energy and quasi-momentum conservation
as illustrated in Fig.~\ref{fig1}. The phase-matched quasi-momenta
$k_0$ occur within the interval: $k_L/2<k_0<k_{zm}$, where
$k_L=Q/2$ and $k_{zm}$ is a point of zero effective mass in the
band structure ($\frac{d^2\varepsilon}{dk^2}\big|_{k_{zm}}=0$,
$0<k_{zm}<k_L$). Those sets of wave vectors are plotted as the
full line in Fig.~\ref{fig3}. This simple procedure produces
results very much in agreement with those obtained by the full
numerical solution. For completeness we present also a calculation
including the effect of interactions. To calculate the band
structure for quasi-momentum states of atoms outside the $k_0$
condensate state, we assume that these are particle-like
excitations, and we solve the following equation:
\begin{eqnarray}\label{popov}
\Big(-\frac{\hbar^2}{2m}\frac{d^2}{dz^2}+V(z)+2\gamma|\Psi_0(z)|^2\Big)
u(z)=\varepsilon u(z),
\end{eqnarray}
sometimes referred to as the Popov approximation to the coupled
Bogoliubov-de Gennes equations for the problem. The mean-field
interaction term $2\gamma|\Psi_0(z)|^2$ depends on $k_0$ and
amounts to less than a five percent correction to $V(z)$. Hence,
we identify $\Psi_0(z)$ and we solve Eq.~(\ref{popov}) for each
quasi-momentum $k_0$ and derive band structures, from which we
extract the phase-matched pair $k_1,k_2$ of final momenta. Sets of
$(k_0,k_1,k_2)$ found with this method are shown by the dashed
curve in Fig.~\ref{fig3}. As expected the interactions only
slightly change the phase-matching condition.

\begin{figure}
\includegraphics[scale=0.89]{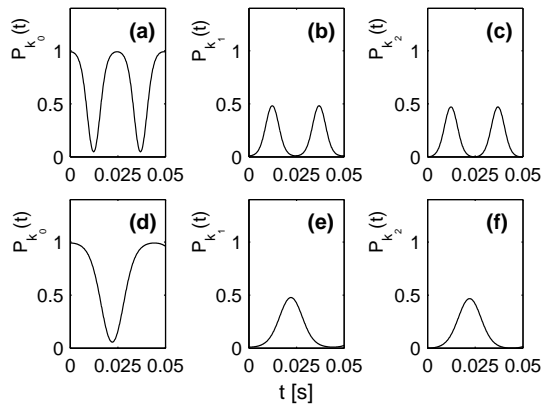}
\caption{The condensate evolution based on the Gross Pitaevskii
equation for sets of phase-matched wave vectors:
(a):~$P_{k_0=0.672k_L}(t)$, (b):~$P_{k_1=1.055k_L}(t)$ and
(c):~$P_{k_2=0.289k_L}(t)$ for $\beta=1/2$ and $N=100000$. The
evolution of the phase-matched components of the condensate with
half the number of atoms ($N=50000$) is shown in
(d):~$P_{k_0=0.672k_L}(t)$, (e):~$P_{k_1=1.047k_L}(t)$ and
(f):~$P_{k_2=0.297k_L}(t)$. Half the oscillation frequency is
observed since the oscillation frequency is proportional to the
number of atoms.} \label{fig4}
\end{figure}

Our original expectations were that the phase-matched, degenerate
FWM could be achieved in a perturbative regime with only few atoms
expelled from the original condensate. When the process is
phase-matched, however, the calculations show an extremely high
conversion efficiency (up to 95\%). Furthermore the populations
$P_{k_i}(t)$, displayed more clearly in Fig.~\ref{fig4}(a)-(c) for
the set of phase-matched wave vectors
$(k_0,k_1,k_2)=(0.672k_L,1.055k_L,0.289k_L)$, show clear
oscillatory behavior. Such Rabi-like oscillations are normally met
in transitions between discrete states, but, as illustrated in
Fig.~\ref{fig1}, increasing $k_1$ (and decreasing $k_2$) lowers
the energy of both states, whereby energy conservation restricts
the coupling to a narrow part of the momentum state continuum, in
which case the dynamics passes to Rabi oscillatory dynamics
\cite{cohen92a}. The expected frequency of the Rabi-oscillations
is proportional to  $N$. Fig.~\ref{fig4}(d)-(f) illustrate
$P_{k_0}(t),P_{k_1}(t),P_{k_2}(t)$ for the phase-matched
$(k_0,k_1,k_2)=(0.672k_L,1.047k_L,0.297k_L)$ from a
Gross-Pitaevskii simulation with half the amount of atoms as
compared with Fig.~\ref{fig4}(a),(b),(c), and, indeed, we observe
approximately half the Rabi-oscillation frequency. Note the
slightly different values of ($k_1,k_2$) due to the effect of the
interactions on the phase-matching condition.

Assuming the rather strict final state selectivity due to energy
and momentum conservation we have performed a number state
analysis in a few mode basis based on the three populated
quasi-momentum states. We have thus written the wave function as
$\Psi=\sum_{n_{k_0},n_{k_1},n_{k_2}}c_{n_{k_0},n_{k_1},n_{k_2}}|n_{k_0},n_{k_1},n_{k_2}>
=\sum_{n}c_n|N-s-2n,s+n,n>$ and calculated the evolution of the
$c_n$ coefficients, starting with $N-s$ atoms in $k_0$ and $s$ in
$k_1$, and evolving under a Hamiltonian that removes pairs of
$k_0$ atoms and creates $k_1,k_2$ pairs. This calculation confirms
the oscillatory behavior between the mean number of atoms in the
$k_0$ state and in the $k_1,k_2$ states. While the numerical
solution of the Gross-Pitaevskii equation shows that even a tiny
seed leads to the same conversion, but delayed with respect to the
results of a larger seed, our number state calculation shows that
it is necessary to seed the condensate by a finite amount to avoid
the decoherence due to strongly n-dependent coupling amplitudes.
With sufficient seed our number state analysis reproduces the
macroscopic conversion and population oscillations observed in the
mean field analysis.

In conclusion the process of phase-matched FWM in a BEC in a
periodic potential can be extremely efficient.  When seeded, up to
95\% of the atoms originally in the condensate with wave vector
$k_0$ can be transferred into the correlated pair of states with
$k_1$ and $k_2$. The system can be used as a source of correlated
atomic clouds where pairs can be easily separated due to the
rather large momentum difference between $k_1$ and $k_2$. Because
of the coupling to a very narrow continuum of states the system
performs Rabi-oscillations between the $k_0$ and the $k_1,k_2$
states, a behavior confirmed by the number state analysis with few
modes.

Potentially better descriptions than our simple Gross-Pitaevskii
equation \cite{modugno04a} have been proposed to describe
transversely confined elongated condensates. Our main mechanism
relies on the non-trivial band structure, but not on its
particular shape, and we therefore believe that our proposal
should remain generally valid. Transverse confinement is, however,
an important issue, and it is a natural extension of our theory to
consider transverse excitations and, more generally, motion in 2D
or 3D lattices, where energy conservation and phase-matching may
lead to a range of interesting solutions.

Theories \cite{wu03a,modugno04a} have proposed and experiments
\cite{burger01a,fallani04a} have shown that condensates moving in
periodic potentials become unstable for certain ranges of
quasi-momenta. These results are linked with the energy and
momentum conserving processes identified in this paper, but they
also involve the detailed properties of the transverse confinement
\cite{modugno04a}. Our calculations assume a lower density of
atoms than in the experiments reported in \cite{burger01a}, and in
this regime we expect that only few atom pairs will be
spontaneously scattered, and that our seeded process will be
dominant.

K.M. Hilligs\o e acknowledges NKT Academy for financial support
and Kevin Donovan for carefully reading the manuscript.

\end{document}